\documentstyle[12pt,epsf]{article}
 \def \bfgr #1{ \mbox {{\boldmath $#1$}}}
 \def \dfrac #1#2 {\displaystyle\frac{#1}{#2}}
\begin{document}
 {\bf \begin{center}
 The Neutron Spin Structure Function from the Deuteron Data\\
 in the Resonance Region
 \end{center} }

 \vskip 5mm

 \begin{center} C.  Ciofi degli Atti,
 L.P.  Kaptari\footnote {\em On leave from
   Bogoliubov Lab.  of Theor.  Phys., Joint Institute for Nuclear Research,
   Dubna, Russia.},
 S.  Scopetta and A.Yu.  Umnikov \\[2mm]
{\em Department of Physics, University of Perugia, and
INFN, Sezione di Perugia,
\\
via A. Pascoli, Perugia, I-06100, Italy.}
\end{center}

 \vskip 5mm

 \begin{abstract} Nuclear effects in the spin-dependent structure function
 $g_1$ of the deuteron are studied in the kinematics of future experiments at
 CEBAF, ($\nu \leq 3~GeV, ~Q^2 \leq 2~GeV^2$).  The magnitude of nuclear
 effects is found to be significantly larger than the one
occurring in deep inelastic scattering
($\nu\to \infty, ~Q^2\to \infty$).  A possibility to measure the
 neutron structure functions in the CEBAF experiments with deuterium is
 analysed.  It is found that disregarding or improperly treating  nuclear
 effects in the region of nucleon resonances would lead to the ``extraction"
 of an unreliable function. A procedure aimed at correctly
extracting the
neutron structure function from the deuterium data is
illustrated
and conclusions about
the experimental study of the $Q^2$ dependence 
of the Gerasimov-Drell-Hearn Sum
 Rule for the neutron are drawn.

 \end{abstract}

 {\bf I}. Recently it has been 
 proposed~\cite{cebaf} at CEBAF to experimentally study
 the spin-dependent structure function (SF)
 of the neutron $ g_1^n$, in a wide interval of 
 energy $\nu$ ( $0.2 - 3~GeV$) and 
 momentum transfers $~Q^2$ ($0.15 - 2~GeV^2$),
 using 
 polarized deuterium and $^3He$ targets. 
 These experiments will shed light on
 a number of
 quantum chromodynamics (QCD) sum rules and 
will help to establish a connection between
 results predicted by low energy theorems ($Q^2\to 0$) and perturbative QCD
 ($Q^2 \gg m^2,$ $m$ being the nucleon mass). 

Of particular interest is
 the $Q^2$ dependence of the Gerasimov-Drell-Hearn (GDH)
 Sum Rule for the neutron.
 
 The GDH Sum Rule, which has been derived 
 in the real photon limit ($Q^2=0$) by Gerasimov~\cite{ger1}
 and Drell and Hearn~\cite{drell}, reads as follows: 
 \begin{eqnarray}
 \frac{m^2}{8\pi^2\alpha}\int\limits_{\nu_{th}}^\infty\frac{d\nu}{\nu}
 \left(\sigma_{1/2}(\nu)-\sigma_{3/2}(\nu)\right )=-\frac{\kappa^2}{4},
 \label{gdh} 
\end{eqnarray}
 where $\kappa$ is the nucleon anomalous magnetic
 moment, $\nu_{th}$ is the threshold energy of the pion photo-production,
 $\sigma_{1/2(3/2)}$ is the absorption cross section for total helicity
 $1/2(3/2)$, $\alpha$ is the fine structure constant.
 The sum rule (\ref{gdh}) can be generalized to the case of electron
 scattering by expressing the helicity cross sections, $\sigma_{1/2(3/2)}$,
 through
 the spin-dependent SF $g_1(x,Q^2)$ and $g_2(x,Q^2)$,
 obtaining~\cite{drech}
 \begin{equation} I_{GDH}(Q^2)=
\frac{2m^2}{Q^2}\int\limits_{0}^{x_{max}} dx
 \left [g_1(x,Q^2) - \frac{4m^2x^2}{Q^2}g_2(x,Q^2)\right ].
  \label{gener}
 \end{equation}
 Since under  usual experimental conditions the second term in
 eq.  (\ref{gener}) is small,
 one expects that
 \begin{equation} I_{GDH}(Q^2=0)
 = -\frac{\kappa^2}{4},
 \label{igdh}
 \end{equation}
 i.e. a negative value for the  first moment
 of the nucleon SF. Since the experimental data in the
  region of the deep inelastic scattering show a positive
 value for the first moment of proton SF~\cite{emc,smc,slac},
 this implies that
 in the $Q^2$-evolution of
 the GDH Sum Rule from the real photon limit  to
 the deep inelastic region, the
 first moment of the proton SF must change its sign.
  The change
 in sign is expected to occur 
in the region of excitation of the
 nucleon resonances
 where neither perturbative QCD nor
 chiral theories are applicable.
  Indeed,
 the momentum transfer
 is already small enough to use the
 perturbative methods of QCD, whereas, at the same time, the
 chiral loops expansion does not
 work because of drastic changes of the helicity structure in the resonance
 region.

 On the other hand side, the measurement of the neutron SF will
 essentially
 contribute to the analysis of the deep inelastic sum rules,
 such as
 the Bjorken sum rule (BSR) and Ellis-Jaffe sum
 rules~(see, for instance ref.~\cite{efremov} and
 references therein). 
 In the deep inelastic limit, $Q^2\to\infty$, the
 BSR connects the difference of the
 first moments of the spin-dependent SF of the proton
 and the neutron with the axial
 constant of the neutron $\beta$-decay:
  \begin{eqnarray} \int\limits_0^1 dx\,\left ( g_1^p(x) - g_1^n(x)\right ) =
 \frac{1}{6}\left( \frac{g_A}{g_V}\right ), \label{bsr} 
\end{eqnarray} where
 $x=Q^2/2m\nu$ is the Bjorken scaling variable.
 The experimental check of eq.~(\ref{bsr}) has already
 begun~\cite{emc,smc,slac} by a measurement of the SF of the proton
 and the neutron, using in the latter case
nuclear targets, viz.
 deuterium and $^3He$.  The experimental data from 
 different groups give slightly different values for
 the BSR.  However, theoretical efforts in computing the $Q^2$-corrections
 recoincile the data with the theoretical prediction (\ref{bsr}),
 leading to the conclusion that the BSR is
 experimentally confirmed
 with an accuracy of about two standard deviations in the
 measured interval of $Q^2$~\cite{efremov}.
  Nevertheless, the 
 investigation of the $Q^2$-evolution 
 of the BSR in a wide interval of $Q^2$
 and the problem of its explanation within QCD
 remain of 
 great interests.
 Beside the GDH and
 BSR sum rules, yelding integral characteristics 
 of the SF, a
 detailed study of the resonance behavior
 of the nucleon SF $g_1(x,Q^2)$ 
 is planned as well.

 All above examples demonstrate that
 the measurement of the spin-dependent neutron SF at CEBAF
 will provide us
 with important new information about 
 the nucleon structure and will help 
 to test a number of  theoretical models and methods.

  Keeping in mind the lessons we have learned
 from the EMC-effect, one might
expect that
nuclear corrections could play an 
 important role in estimating the neutron SF from
 the combined nuclear and proton data~\cite{amb,mt}. In the
 region of finite $Q^2\sim m^2$ and
 $\nu\sim m$, nuclear corrections are
 much more important than in
 the deep inelastic limit~\cite{cs}.
 In this paper the role of nuclear structure effects
 in electron-deuteron scattering in the
  resonance region will be discussed, paying
special attention to the procedure of the extraction
  of the neutron SF from the deuteron data
 in the kinematics of future
  experiments at CEBAF.

 {\bf II}. 
  The nucleon contribution to the deuteron structure functions
 is usually calculated by weighting the amplitude
 of  electron scattering on the nucleon with the wave function of nucleon
 in the deuteron (for recent developments see
 e.g.~\cite{ourphysrev,ourphyslet,mt2,kmpw}
 and references therein).
 For the spin-dependent SF the most important effects 
 are the Fermi motion  and the depolarizing effect of
 the D-wave.
 Additional effects, such as off-mass-shell effects 
 or nucleon deformation,
 are found to be small~\cite{mst,offmass}.
 For finite values of $Q^2$ and $\nu$, the 
 deuteron SF $g_1^D(x,Q^2)$ reads as follows
  \begin{eqnarray} 
g_1^D(x,Q^2)&&\!\!\!\!\!\!\!\!=
\int \frac{d^3{\bfgr k}}{(2\pi)^3}
\frac{m\nu}{kq}
 g_1^N\left (x^*,Q^2 \right )
\left ( 1+\frac{\xi(x,Q^2) k_3}{m}\right )
  \left ( \Psi_D^{M+}({\bfgr k}) S_z \Psi_D^M({\bfgr k}) \right)_{M=1}
 \label{gd}\\[2mm]
 &&\!\!\!\!\!\!\!\!=\int\limits_{y_{min}(x,Q^2)}^{y_{max}(x,Q^2)}
\frac{dy}{y} g_1^N(x/y,Q^2)
 \vec f_D(y,\xi(x,Q^2)),
 \label{conv} 
\end{eqnarray}
 where $g_1^N=(g_1^p+g_1^n)/2$ is the isoscalar nucleon SF and
 $\Psi_D^M({\bfgr k})$\ the deuteron wave function  with spin
 projection $M$. In the rest-frame
of the deuteron, with ${\bfgr q}$ opposite the z-axis,
kinematical variables are defined as:
\begin{eqnarray}
&& kq = \nu (k_0 + \xi(x,Q^2)k_3),\quad k_0 = m+\epsilon_D- {\bfgr k}^2/2m,\\
&&\!\!\!\!\!\!\!\!\!\!\!\!\!\!
\xi \equiv q_3/\nu= |{\bfgr q}|/\nu = \sqrt{1+4m^2x^2/Q^2},
\quad Q^2 \equiv -q^2,\quad x^* = Q^2/2kq,
\end{eqnarray}
where $\epsilon_D=-2.2246~MeV$ is the deuteron binding energy.
The limits ${y_{min (max)}(x,Q^2)}$ are defined
to provide an integration over the physical region of
momentum in (\ref{gd}) and to take into account the 
pion production
threshold  in the virtual photon-virtual nucleon
scattering\footnote{For $x$ not too close to the limit
of single-nucleon kinematics, $x\to 1$,
the quasi elastic contribution can be disregarded}.
Since both ${y_{min}(x,Q^2)}$ and  ${y_{max}(x,Q^2)}$
are solutions of a transcendent equation,
explicit expressions for them cannot be given. However,
in our numeric calculations they are 
accurately taken into account.

Eqs. (\ref{gd})-(\ref{conv}) have the correct limit in
the deep inelastic kinematics
($Q^2\to \infty, \quad \nu \to \infty$).
In this case: $\xi(x,Q^2) \to 1, \quad y_{min} 
\to x, \quad y_{max}\to M_D/m$, and the usual convolution
 formula for the deuteron SF~\cite{ourphysrev,kmpw} is recovered: 
 \begin{eqnarray} 
g_1^D(x,Q^2)=\int\limits_{x}^{M_D/m}\frac{dy}{y} g_1^N(x/y,Q^2)
 \vec f_D(y).
 \label{convdis} 
\end{eqnarray}
Equation (\ref{convdis}) defines spin-dependent  
``effective distribution
of the nucleons", $\vec f_D$, which describes the bulk of the nuclear 
effects in  $g_1^D$. The main features of the distribution function, 
$\vec f_D(y)$, are a sharp maximum at $y = 1+\epsilon_D/2m\approx 
0.999$ and a normalization given by $(1-3/2P_D)$
 ($P_D$ being the weight of the D-wave in the deuteron). 
As a result in the region of medium values of $x\sim 0.2-0.6$ 
the deuteron SF $g_1^D(x)$ is slightly
 suppressed  by a depolarization
factor, $(1-3/2P_D)\times g_1^N(x)$, 
compared to the free nucleon SF. However,
the magnitude of this suppression is small ($\sim 1\%$) and this is
why it is phenomenologically acceptable to extract the neutron SF
from the deuteron and proton data by making use of the following
approximate formula:
 \begin{equation} g_1^D(x,Q^2)\approx \left
 (1-\frac{3}{2}P_D\right )( g_1^n(x,Q^2) + g_1^p(x,Q^2))/2. 
 \label{extract}
 \end{equation}
In addition, when integrated over $x$, eqs. (\ref{convdis}) and
(\ref{extract}) give exactly the same result ($\Gamma = \int dx g_1(x)$), i.e.
 \begin{equation}
  \Gamma_D(Q^2) = \left(1-\frac{3}{2}P_D\right )
(\Gamma_n(Q^2) + \Gamma_p(Q^2))/2,
 \label{gamma}
 \end{equation}
which allows to define {\em exactly} the integral of the neutron SF
$\Gamma_n$ from the deuteron and proton integrals,
without solving (\ref{convdis}).

Eqs. (\ref{gd})-(\ref{conv})
at finite values of $Q^2$ and $\nu$ are more sophisticated than 
the corresponding equations in the deep inelastic limit.
 In particular, they do not represent
 a ``convolution formula" in the 
 usual sense, since the effective distribution function  $\vec f_D$
 and the integration limits
 are also functions of $x$. This circumstance immediately leads to the
 conclusion that, in principle, when integrals
 of the SF are considered, the effective distribution
 can not be integrated out to get the factor similar to $(1-3/2P_D)$
in (\ref{gamma}).
 Another interesting feature of formulae (\ref{gd})-(\ref{conv})
 is the $Q^2$-dependence of  $\vec f_D$ and 
${y_{min,(max)}(x,Q^2)}$. If we again limit ourselves to
the discussion of the integrals of SF, one concludes
that the $Q^2$-dependence
of such an integral is governed by both the QCD-evolution of the nucleon
SF and the kinematical $Q^2$-dependence of the effective distribution
of nucleons. 

Thus,  we have established that in the non-asymptotic regime, equation
 (\ref{gamma}), in principle, does not hold. Furthermore, it is not clear 
 whether an equation similar to (\ref{extract}) could be 
 applied in this region. Indeed, we are discussing the kinematical conditions 
 pertaining to 
nucleon resonances, where the ``elementary" nucleon SF
 explicitly exhibits Breit-Wigner resonance structures  corresponding
 to the excitations of the nucleon by the photon
 and one expects that the Fermi motion and binding of nucleons will result in 
 a shift and a smearing of the resonance structures. 
 However, one can  hope that actual effects  will be quantitatively 
 small so that eq. similar to
 (\ref{gamma}) and (\ref{extract}) could
  phenomenologically still be valid.

 {\bf III}.  In our numerical estimates we use
 a reliable parametrization of the proton and neutron SF given
 by Burkert~\cite{burkert},
 which
 takes into account several nucleon excitations and
 provides  a reasonable
 description of the available nucleon data in the resonance region.
 Using the Bonn potential model for
the deuteron wave function~\cite{bonn}, we carry out 
a realistic calculation of the deuteron
 SF, $g_1^D(x,Q^2)$ in the region of nucleon resonances. 
  
\begin{figure}[ht] \epsfxsize 4in

 \hspace*{-0.5cm} \epsfbox{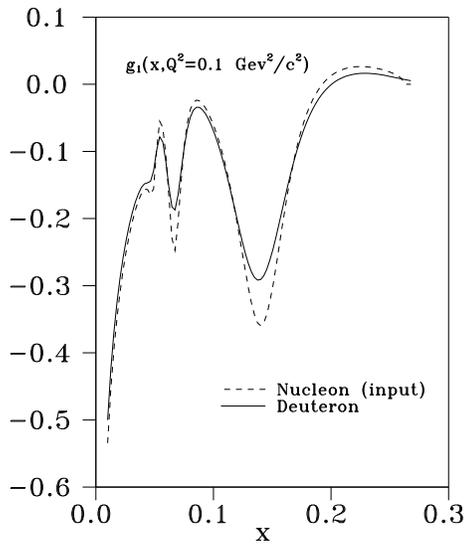}

 \vspace*{-14cm} \epsfxsize 4in \hspace*{7cm} \hfill \epsfbox{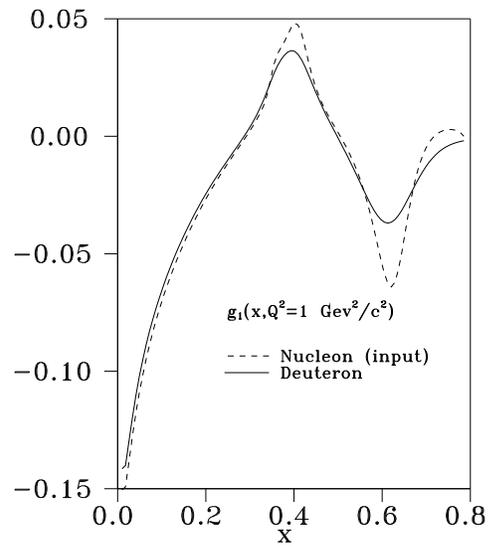}

 \vskip 3.5cm \caption{
 C. Ciofi degli Atti et al, 
The Neutron Spin Structure Function from\ldots}
 \label{gdh-sf}
 \end{figure}

   In Fig. \ref{gdh-sf} the results 
 of the calculation of the deuteron SF,
 $g_1^D(x,Q^2)$ at $Q^2 = 0.1$~GeV and $1.0$~GeV, 
are compared with the input of the calculation, i.e. the isoscalar
 nucleon SF, $g_1^N(x,Q^2)$.
It can be seen that
the role of nuclear effects in the 
resonance region is much larger (up to $\sim 50\%$ in
the maxima of the resonances),
than in the deep inelastic regime ($\sim 7-9\%$, depending upon
  the models~\cite{ourphysrev,ourphyslet,mt2,kmpw}, with resulting
 $\sim 6-7 \%$ from the
 depolarization factor  $(1-3/2P_D)$ and  $\sim 1-2\%$ from the
 binding effects and Fermi motion).
 Such a drastic effect is a consequence of the presence of the
 narrow resonance peaks in the nucleon SF.

\begin{figure}[ht] \epsfxsize 4in \hspace*{-0.5cm}\epsfbox{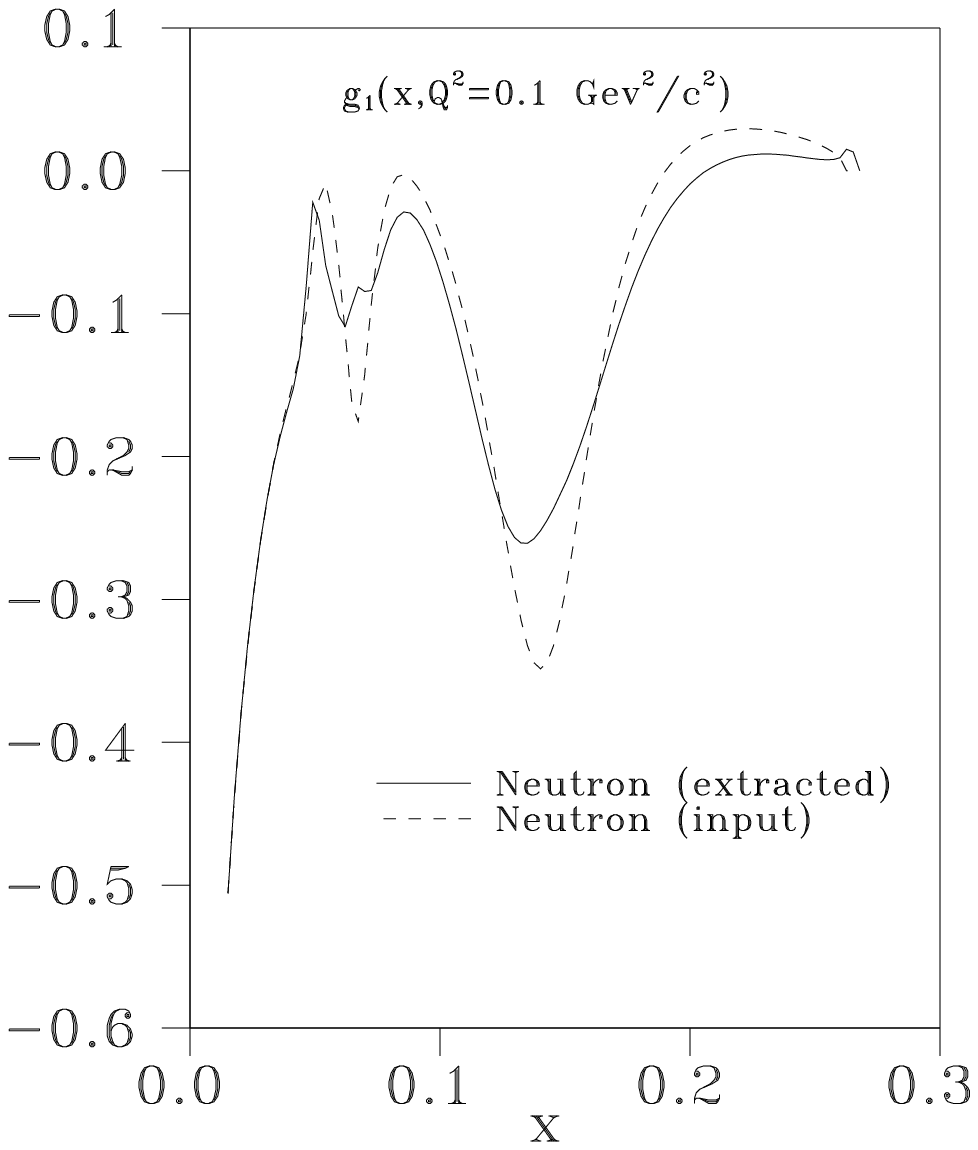}
 \vspace*{-14.2cm} \hspace*{6cm} \epsfxsize 4in \hfill \epsfbox{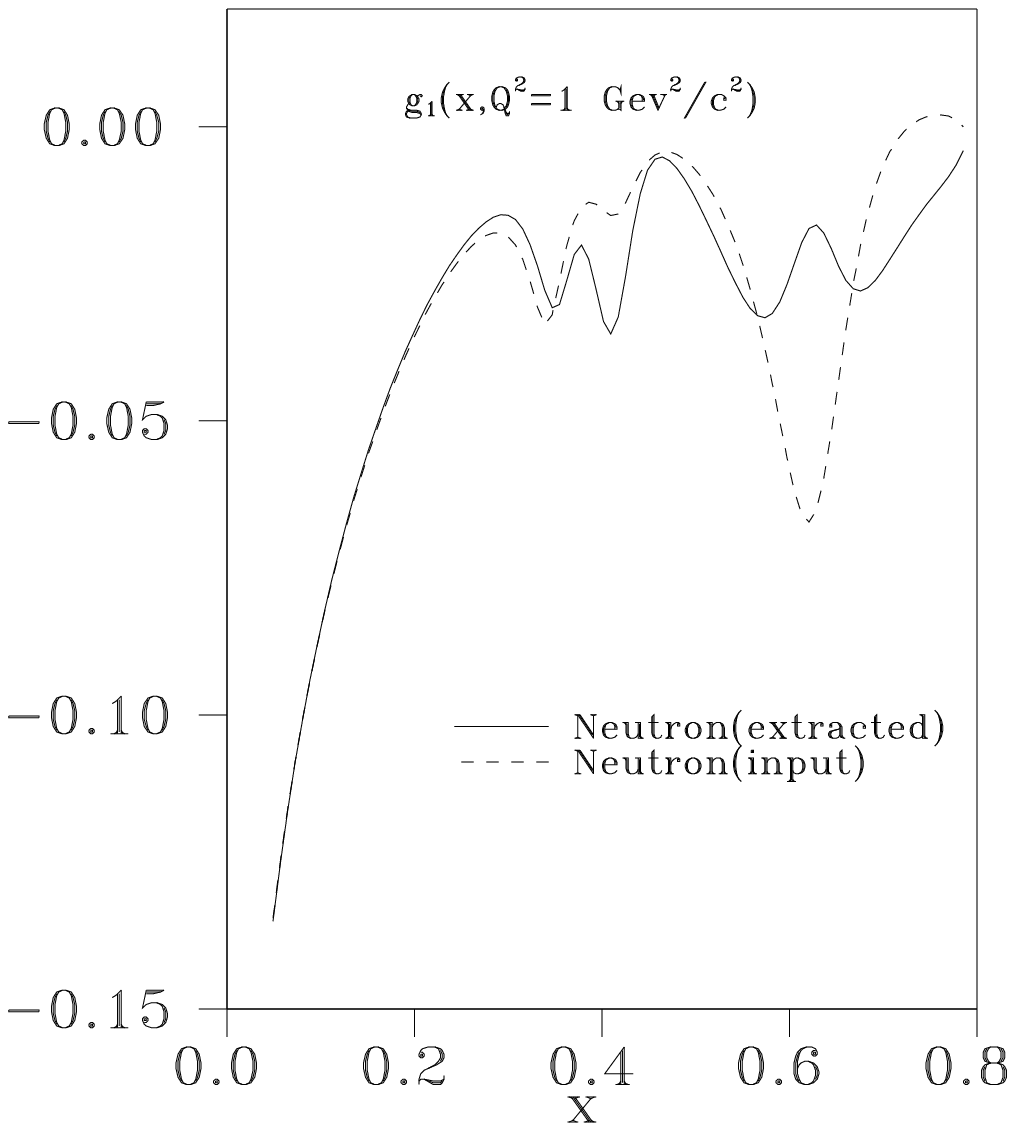}
 \vskip 3.5cm \caption{  C. Ciofi degli Atti et al, 
The Neutron Spin Structure Function from\ldots
  }

\label{gdh-ntr} 

 \end{figure}

 Fig. \ref{gdh-ntr} shows the results of extraction of the neutron
 SF from the deuteron and proton data by 
using the approximate formula (\ref{extract})
 which, we believe to give an upper limit of the possible errors
 in this extraction.
   To emphasize
 the role of nuclear effects in the region of finite $Q^2$, 
 the extracted neutron SF is compared
 with the original (input in the calculation) parametrization of the 
 neutron SF. The use of the
 approximate formula (\ref{extract}) appears to be in some regions
 completely unreliable. 
 This can be easily understood as follows:
 the proton and neutron SF 
 have
 similar behavior in the resonance region, in that the positions of the
 nucleon resonances are the same for both of them,
 whereas the resonances in the 
 resulting deuteron SF are smeared and
 shifted, compared to the isoscalar SF.
 Therefore,
 subtraction the proton SF from the deuteron one,
 in the maximum of the former,
 can result in a minimum for the neutron SF, instead of a maximum.
 The conclusion of our analysis is 
 that nuclear effects in the resonance region are
 very specific and the approximate formula (\ref{extract})
  does not work  even for crude extraction
 of the neutron SF. 
  Obviously, another 
 method of extracting the neutron SF should be used.

 In ref.~\cite{unfold} a rigorous
 method of solving
 eq. (\ref{convdis}) for  the unknown neutron SF has been proposed and
 applied in the deep inelastic region. 
 It has been shown that this method,
which works for both 
spin-independent and
spin-dependent SF, 
 in principle allows one to
 extract the neutron SF exactly, requiring
 only the analyticity of
 SF. It can also be applied by a minor modification
to the extraction of the SF at finite $Q^2$, which is our present
aim.

  The basic idea is to replace the integral equation
 (\ref{conv}) by a set of linear algebraic equations, $K{\cal G}_N ={\cal
 G}_D$, where $K$ is a square  matrix
(depending upon the deuteron model),
 ${\cal G}_D$ is a vector
 of the experimentally known deuteron SF and ${\cal G}_N$ is 
a vector of unknown
 solution. 
 Changing the integration variable
 in (\ref{conv}), $\tau =
 x/y$, we get
 \begin{eqnarray} 
g_1^D(x,Q^2)=\int\limits_{\tau_{min}(x,Q^2)}^{\tau_{max}(x,Q^2)}
d\tau g_1^N(\tau,Q^2)
\frac{1}{\tau} \vec f_D(x/\tau,\xi(x,Q^2)),
 \label{conv2} 
\end{eqnarray}
where $\tau_{min}(x,Q^2)= x/y_{max}(x,Q^2)$,
$\tau_{max}(x,Q^2)=x_{max}(Q^2)/y_{min}(x,Q^2)$ and $x_{max}(Q^2)$ is 
defined by the pion production threshold in virtual photon-nucleon
scattering. Let us assume that the deuteron SF has been measured
 experimentally in the
 interval $(x_1,x_2)$ and a reasonable parametrization for the SF is found in
 this interval. 
 Then, dividing both intervals $(x_1,x_2)$ and $(\tau_{min}, 
\tau_{max})$
 into $N$ small parts, one may write:
  \begin{eqnarray} && g_1^D(x_i,Q^2)\approx
 \sum_{j=1}^N g_1^N(\tilde\tau_j,Q^2)\int\limits_{\tau_j}^{\tau_{j+1}}
\frac{1}{\tau} \vec f_D(x_i/\tau,Q^2) d\tau, \quad i=1\ldots N,
\label{mat}
\end{eqnarray}
where $\tilde\tau_j = \tau_{min}+h(j-1/2)$ and 
 $h=(\tau_{max}-\tau_{min})/N$. Equation (\ref{mat})
 is already explicitly of the form 
 ${\cal
 G}_D=K{\cal G}_N$, therefore the usual linear algebra methods can be
applied to solve it.

Note that the range of variation of
$\tau$ is larger than the one for $x$.
Therefore, in principle, the SF of the deuteron experimentally  
known in the interval $(x_1,x_2)$ contains
information about neutron SF in wider interval (for example, in
deep inelastic regime $\tau_{min}\approx x/2$ and
$\tau_{max}=1$). However extracting information beyond the interval
 $\tilde \tau_{min}=x_1$ to
$\tilde\tau_{max}=x_2$ is almost impossible in view of the structure
of the kernel of eq.~(\ref{conv2}) and
the kinematical condition of 
planned experimental data~\cite{unfold}. We have to redefine 
the kernel  of eq.~(\ref{conv2}) to incorporate new limits
of integration  $\tilde \tau_{min}=x_1$ and
$\tilde\tau_{max}=x_2$~\cite{unfold}.

The procedure of solving  eqs.~(\ref{conv}) in the kinematical region
of finite $Q^2$ and $\nu$ will be presented elsewhere
in details; here we only stress that
the method works with a good accuracy.  
To check it, we calculated the deuteron
 SF by formula (\ref{conv}) with the nucleon SF
 $g_1^N(x,Q^2)$ from ref.~\cite{burkert} and the deuteron wave function of
 the Bonn potential~\cite{bonn}.  Then the obtained $g_1^D(x,Q^2)$ 
 has been used as
 ``experimental'' data to calculate the vector 
${\cal G}_D$ in (\ref{mat}); the
 matrix $K$ has been calculated by 
using the same deuteron wave function.
Equation (\ref{mat}) 
 has been solved numerically for various ``experimental''
 situations (changing the ``measured''
 interval $(x_1,x_2)$, for different $Q^2$,
 etc.).  The obtained solution, i.e.
 the extracted neutron SF, has been compared
 point by point with the input to the calculation of
 $g_1^D$.  We found that  method is stable and allows one to unfold
 the neutron SF with
 errors not larger than $10^{-4}$,
 which is much smaller than the expected
 experimental errors
 (note, that in the present paper we discuss only the spin dependent SF,
 nevertheless all results and conclusions are valid for unpolarized structure
 function $F_2(x,Q^2)$ as well.) 

 {\bf IV}.  In this section we discuss
 the role of nuclear corrections in the 
 analysis of the integrals of the SF, such as the GDH  Sum Rule.

 A  very important observation has been made
 in the deep inelastic limit, namely 
 the {\em exact} formula (\ref{convdis}) and the {\em approximate} formula
 (\ref{extract})  give the same result for the integral of the neutron
  structure function, $g_1^n(x,Q^2 \gg m^2)$ (see eq.~(\ref{gamma})).
 The applicability of the approximate formula in the deep
inelastic region is based on the conservation
of the norm of the distribution $\vec f(y)$  by the convolution formula
(\ref{convdis}).
This circumstance can not be immediately
extended to the  case of the resonanse region, since: (i)
the covolution is broken in eq. (\ref{conv}) and (ii) 
the normalization of the function $\vec f(y,x,Q^2)$ is different from
one of $\vec f(y)$.  The integral of the distribution $\vec f(y,x,Q^2)$
 represents the
``effective number" of nucleons ``seen" 
by the virtual photon in the process when
the virtual photon is absorbed by the nucleon
and at least one pion is produced
in the final state (it   is less than 1 at low $Q^2$
and $x \to x_{max}$).

However, surprisingly enough,  
the use of the formula~(\ref{gamma}) in the
resonance region gives results numerically very 
close to the integration of the exact equation~(\ref{conv}).  
This is a consequence of the smallness of the effects breaking 
the convolution in eq.~(\ref{conv}).
These effects can be accounted for by a new equation:
 \begin{equation}
  \Gamma_D(Q^2) = \left(1-\frac{3}{2}P_D\right )N_{eff}(Q^2)
(\Gamma_n(Q^2) + \Gamma_p(Q^2))/2
 \label{gamma1}
 \end{equation}
Eq.~(\ref{gamma1})
 and the integral of eq.~(\ref{conv})
represent the definition of the "effective number" $N_{eff}(Q^2)$;
the latter depends upon the form of the nucleon SF
$g_1^N$, and, since this is expected to strongly oscillates
(see Fig.~1), even the sign of the correction can vary.
For instance we obtain using the SF from~\cite{burkert}, 
 \begin{equation}
N_{eff}(Q^2=0.1~{\rm GeV^2})= 1.02, \quad
 N_{eff}(Q^2=1.0~{\rm GeV^2})= 0.997,
 \label{neffnum}
 \end{equation}
i.e. a rather small effect ($+2\%$ and $-0.3\%$ correspondly).
Therefore eq.~(\ref{gamma1}) appears to be
a reliable one for estimating
the integrals of SF: setting  $n_{eff}(Q^2)=1$ does not lead to 
errors larger than $3\%$ for $Q^2 = 0.1-2.0$~GeV$^2$.

 {\bf V}.  In conclusion, we have shown that the effects of nuclear
 structure  in the extraction of the neutron SF in the
 resonance region 
are much more important  than in the deep inelastic scattering.
  We have explained how the correct neutron SF can be firmly 
 extracted from the combined deuteron and proton data.
 At the same time, we have found that the integrals of the SF,
 such as the GDH  Sum Rule,
 can be estimated with accuracy better
 than 3\% by the simple formula (\ref{gamma}) which is also valid   in 
 deep inelastic region.

 {\bf VI}.  We would like to express our gratitude to V. Burkert,
S.B. Gerasimov and O. Teryaev for enlightining discussions.
 One of us (L.K.) thanks the INFN Sezione di Perugia
 for hospitality during his visit, when most of this work was
 completed.

\newpage
\centerline{\Large \bf  Figure captions:}

\vskip 2cm

Figure 1. The spin dependent structure
 functions $g_1(x,Q^2)$ for two values of $Q^2$.  The deuteron SF (solid line)
 is compared with the isoscalar nucleon SF (dotted line) used as input in
 the
 calculation of eq.~(\protect\ref{conv}).

\vskip 1cm

Figure 2.  The neutron SF (solid
 line) extracted  by the approximate formula (eq.~(\protect\ref{extract}))
compared with the
 original parametrization (dashed line) used in the convolution
 formula~(\protect\ref{conv}).

\newpage

  \newpage

 \end{document}